\documentclass{acm_proc_article-sp}
\usepackage{rotating}

\begin{document}
%
\crdata{0-12345-67-8/90/01}  

\title{DNA Segmentation as A Model Selection Process
\titlenote{
}}
%
%

\numberofauthors{1}
%

\author{
%
\alignauthor Wentian Li  \\
       \affaddr{Laboratory of Statistical Genetics}\\
       \affaddr{The Rockefeller University, Box 192}\\
       \affaddr{New York, NY 10021, USA}\\
       \email{wli@linkage.rockefeller.edu}
       \email{http://linkage.rockefeller.edu/wli/}
}
\date{January 04,  2001}
\maketitle

\begin{abstract}
Previous divide-and-conquer segmentation analyses of DNA sequences
do not provide a satisfactory stopping criterion for the recursion.
This paper proposes that segmentation be considered as a model 
selection process. Using the tools in model selection, a limit
for the stopping criterion on the relaxed end can be determined. 
The Bayesian information criterion, in particular, provides a much more 
stringent stopping criterion than what is currently used.  Such 
a stringent criterion can be used to delineate larger DNA domains. 
A relationship between the stopping criterion and the average domain 
size is empirically determined, which may aid in the determination of 
isochore borders.
\end{abstract}

\section{Introduction}

A typical DNA sequence is not homogeneous. There are local
regions with contrast:  C+G rich versus C+G poor; protein-coding
regions with a strong signal of periodicity-three versus
non-coding regions lacking this periodicity;  high densities
of 5'-CG-3' dinucleotides (CpG island) versus low density of
this dinucleotide; etc. Finding the exact border between these
regions is an important task in DNA sequence analysis.  It is
a common practice to use a moving window to visually monitor the
variation of the quantity of interest (e.g. C+G density) along the
sequence, and the border is determined in an {\sl ad hoc} way.
With the sequence information, it is actually possible to determine
the border exactly by certain mathematical criterion.

These mathematical approaches to delineate regional homogeneous 
domains are known as ``segmentation" \cite{seg_li}, ``partitioning", 
or ``change-point analysis"
\cite{chpt_brod,chpt_car,chpt_chen} in different fields ranging
from image processing to statistics. There are segmentation methods
that require guessing the number of homogeneous regions.
There are also segmentation methods that require specification
of the number of types of domains (e.g.  C+G rich and C+G poor
represent two types of domains, whereas C+G high, intermediate, 
and low specify three types). Segmentation analysis of DNA 
sequences can be found in  \cite{elton,churchill,braun,ramen}.

One particularly attractive segmentation method is a 
divide-and-conquer approach \cite{ber96,ber99} (similar 
recursion processes are also discussed in statistics and machine 
learning under the names of ``classification and regression tree"
\cite{breiman}, ``recursive partitioning" \cite{zhang}, ``decision
tree induction" \cite{quinlan1,quinlan2}, etc). 
The DNA sequence is first segmented into two subsequences so that 
base compositions on two sides of the partition are maximized. Then, 
the same procedure is carried out on both the left and the right 
subsequences; and then on the sub-subsequences, etc. Eventually, 
either the size of a subsequence is too small to segment, or 
the difference between the left and right subsequences is not big 
enough to be worth further segmentation. Recursive segmentation 
offers the following advantages: there is no need to specify the number 
of homogeneous domains beforehand;  the number of types of domains 
need not to be specified (it is implied in the stopping criterion); 
there is no constraint on the size distribution of the domains
(such a constraint exists in hidden-Markov-model-based segmentations);
and the computation is efficient.

This paper addresses one of the disadvantages of this segmentation:
the stopping criterion of the recursion. Another disadvantage of
this approach -- the fact that the solution is a local maximum with 
no guarantee of the global maximum being obtained -- is not 
addressed here. In principle, 
one can set any stopping criterion, leading to domains of any
sizes. In the hypothesis testing framework of statistics, whether
a test is ``significant" or not (corresponding to a continuation,
or a termination, of the recursion in our case) is decided by a pre-set 
``significance level". Usually, the significance level  can be 
0.05, 0.01, or 0.001. These levels are arbitrary and will not
guarantee objectivity \cite{berger}.

We provide a stopping criterion based on the framework of model
selection (for a detailed discussion of the hypothesis testing framework 
versus the model selection framework, see \cite{burnham}). This new
stopping criterion offers at a minimum condition for the recursive 
segmentation to continue. On the other hand, in the hypothesis
testing framework, no such minimum condition exists; for example,
the 0.06 significance level is weaker than the 0.05 level, and
0.1 is even weaker than 0.06, etc. In the model selection framework, 
there are two different guiding principles.  The first is to choose 
a model that most closely approximates the {\sl true model}. 
The second is to find the {\sl true model} among a list of
candidate models.  The first principle leads to a technique of Akaike 
Information Criterion (AIC), and the second leads to the technique 
of Bayesian Information Criterion (BIC). We will show that BIC-based 
stopping criterion for segmentation is practically more useful.

\begin{figure*}
\begin{center}
  \begin{turn}{-90}
  \epsfig{file=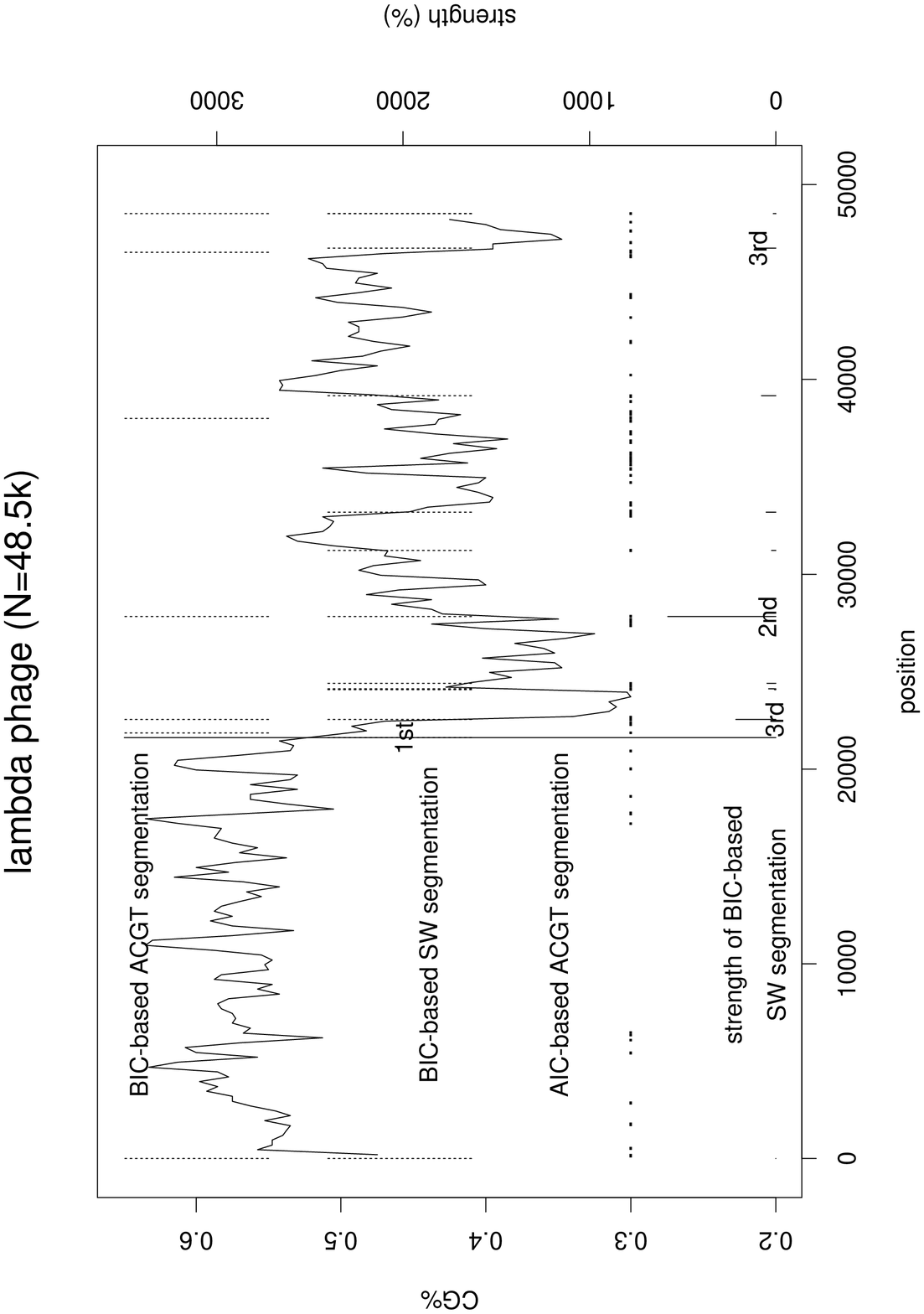, height=15.5cm, width=6.5cm}
  \end{turn}
\end{center}
\caption{
Lambda ($\lambda$) bacteriophage sequence.
}
\label{fig1}
\end{figure*}

\section{Methods}

\subsection{
The divide-and-conquer segmentation in its original formulation}

The original publication of this divide-and-conquer segmentation 
method is called ``entropic segmentation" \cite{ber96,ber99}, 
because the quantities used in determining the partition point 
are based on entropy, a statistical physics concept. The entropy
of a sequence with  length $N$ and number of bases (b) 
$\{ N_\alpha \}$ ($\alpha=a,c,g,t$) is calculated as
( $\hat{}$ means that the quantity is estimated from the data)
\begin{equation}
\hat{E}(\{ N_\alpha\}) =
\sum_{\alpha=a,c,g,t} \frac{N_\alpha}{N} \log \frac{N_\alpha}{N}.
\end{equation}
Given a partition point $i$ ($1 < i < N$), an entropy-based
quantity called Jensen-Shannon distance (divergence) \cite{lin}
is defined as
\begin{equation}
\hat{D}_{JS} = \hat{E}(\{ N_\alpha\}) -\frac{i}{N} \hat{E}(\{ N_{\alpha,l} \})
-\frac{N-i}{N} \hat{E}(\{ N_{\alpha,r} \})
\end{equation}
where $\{ N_{\alpha,l} \}$ and $\{ N_{\alpha,r} \}$ are the base
counts of the left (from position 1 to $i$) and the right
(from position $i+1$ to position $N$) subsequences
(with $\sum_\alpha N_{\alpha,l}=i$,  $\sum_\alpha N_{\alpha,r} = N-i$).
The partition point $i$ is chosen to maximize $\hat{D}_{JS}$.

\subsection{The divide-and-conquer segmentation as a likelihood ratio test}

In fact, the above entropic description can be cast into a
hypothesis testing framework -- the likelihood ratio test \cite{edwards}.
Likelihood is simply the probability of observing the data, given
a model, with emphasis on the functional dependence on the model
parameter (in other words, the normalization coefficient is not
needed). To test whether a model is ``significant", the likelihood
under the model ($L_2$) is calculated,  and maximized over all
possible parameters ($\hat{L_2}$). A similar calculation is carried
out on the null model ($L_1$ and $\hat{L_1}$). If the null model is
the correct model of the data, and if the null model is nested
in the alternative model, it can be shown that in the
large sample size limit  \cite{cox}: 
\begin{equation}
 2 \log \frac{\hat{L_2}}{\hat{L_1}} \sim \chi^2_{df=K_2-K_1}
\end{equation}
where $\chi^2_{df}$ is the chi-squared distribution with degrees
of freedom $df$ (i.e. sum of $df$ terms of squared unit normal
distribution), $K_2$ and $K_1$ are the number of free parameters
in maximizing $L_2$ and $L_1$.

In our divide-and-conquer segmentation, $L_1$ is the likelihood
assuming the sequence being a random sequence, and $L_2$ is
the likelihood assuming two random subsequences:
\begin{eqnarray}
L_1 (\{ p_\alpha \}) &= & \prod_\alpha p_\alpha^{N_\alpha}, \nonumber \\
L_2(\{ p_{\alpha, l} \}, \{ p_{\alpha,r} \}, i) & = &
\prod_{\alpha} p_{\alpha,l}^{N_{\alpha,l} }
\prod_{\alpha} p_{\alpha,r}^{N_{\alpha,r} }
\end{eqnarray}
where $\{ p_\alpha \}$ ($\alpha=a,c,g,t$) is the base
composition of the whole sequence (here these are free
parameters in the model to be estimated), $\{ p_{\alpha,l} \}$
and $\{ p_{\alpha,r} \}$ are the similar base compositions
of the left and right subsequences. The maximum likelihood
estimation of a base composition is simply the percentage
of the base: $\hat{p}_\alpha= N_\alpha/N$. It can easily be
shown that $2 \log(\hat{L_2}/\hat{L_1})$ is the same as
$2N \hat{D}_{JS}$. The number of parameters in the two models
are $K_2=7$ (the partition point $i$ is also
a free parameter) and $K_1=3$. So $2N \hat{D}_{JS}$ under the null
hypothesis should obey the $\chi^2_{df=4}$ distribution (the
same conclusion was reached before, see \cite{ber00} and
(I Grosse, et al. in preparation),  only the $df$ used there
is 3, instead 4).

\begin{figure*}
\begin{center}
  \begin{turn}{-90}
  \epsfig{file=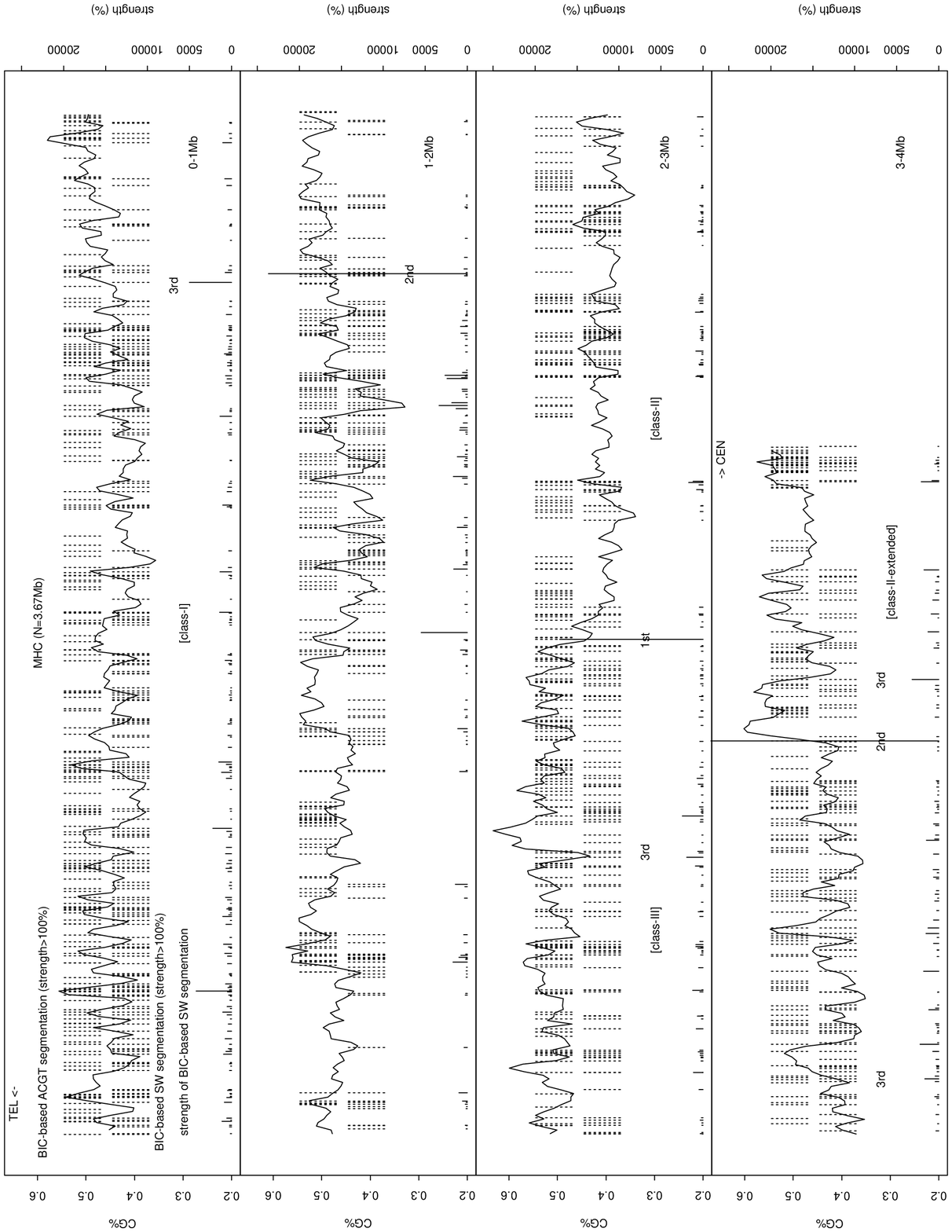, height=16cm, width=17cm}
  \end{turn}
\end{center}
\caption{
Human major histocompatibility complex (MHC) sequence.
}
\label{fig2}
\end{figure*}

\begin{figure*}
\begin{center}
  \begin{turn}{-90}
  \epsfig{file=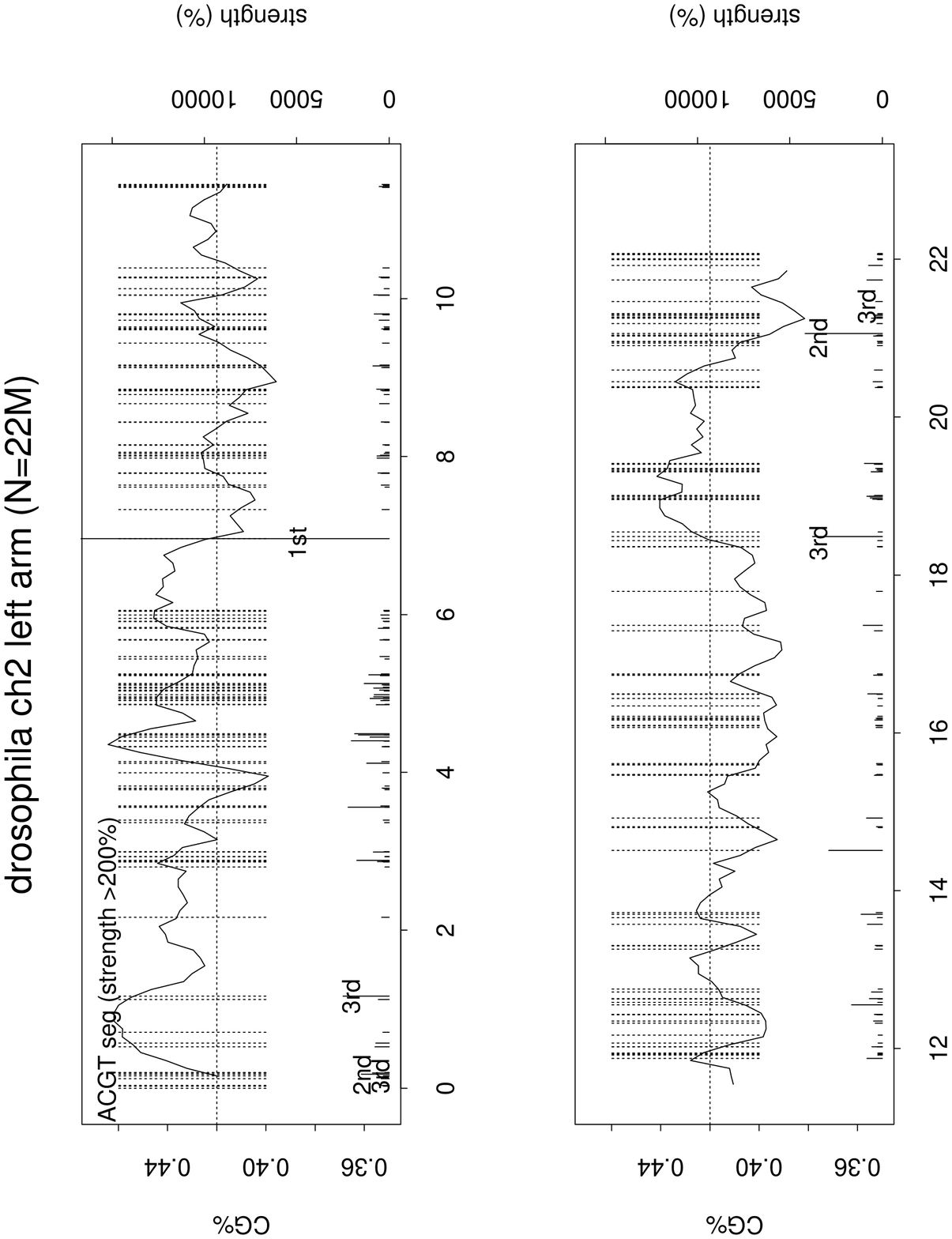, height=16cm, width=7cm}
  \end{turn}
\end{center}
\caption{
Left arm of Drosophila melanogaster chromosome 2
}
\label{fig3}
\end{figure*}

\subsection{The divide-and-conquer segmentation as a model selection}

There are many shortcomings in the hypothesis testing framework
\cite{burnham}. The purpose of a test is to see how bad a description
of the data $L_1$ is, not how good a description $L_2$ is.
For many circumstances, it is not really what we are
interested in. In the model selection framework, we directly address
the ``merit" of a model. One measure of such a ``merit" is whether
the model is close (a better approximation) to the {\sl true model}.
The closeness is measured by the Kullback-Leibler distance (divergence)
\cite{kullback}, and the Akaike Information Criterion (AIC) is
one approximation of this distance (with the constant term
removed, and multiplied by a factor of 2) \cite{aic}:
\begin{equation}
\label{eq_aic}
AIC = -2 \log (\hat{L}) + 2K + O( \frac{1}{N}).
\end{equation}
where $\hat{L}$ is the maximized  likelihood of the model,
$K$ is the number of free parameters in the model. A model
with the lowest AIC is closest to the true model, thus the
best approximating model.

Another ``merit" of a model is how the data increases
the probability of the model (only in Bayesian statistics is it
possible to extend the concept of probability to the model and
its parameters). The factor between the prior and posterior
probability of the model is the ``integrated likelihood" \cite{raftery}.
An asymptotic approximation of minus-twice the logarithm of
the integrated likelihood is the Bayesian Information Criterion (BIC)
\cite{bic}:
\begin{equation}
\label{eq_bic}
BIC = -2 \log (\hat{L}) + \log(N) K + O(1) + O(\frac{1}{\sqrt{N}})
+O(\frac{1}{N}),
\end{equation}
where $N$ is the sample size. A model with the lowest BIC
has the largest integrated likelihood, and this translates to
the largest posterior probability if all models have the
same prior probability. Note that AIC emphasizes an approximation
of the true model, and BIC emphasizes the selection of the true
model from the space of all models.  The high-order terms in
AIC are discussed in \cite{aicc1,aicc2}, and the derivation of
BIC can be found in \cite{raftery}. It can easily be seen that
if $\log(N) > 2 $ (or $N > 7.389 $), the penalty on the number
of model parameters (i.e. the second term in Eq.(\ref{eq_aic})
and Eq.(\ref{eq_bic}))
in BIC is more severe than that in AIC.
As a result, BIC tends to select simpler models than AIC.

When the segmentation is viewed as a model selection process,
the model before the segmentation describes the sequence as a
random sequence, whereas that after the segmentation describes
it as two random subsequences. Since AIC/BIC must decrease
for the segmentation to continue, it can be shown that they lead
to the two stopping criteria as follows:
\begin{eqnarray}
\mbox{AIC-based}  & & \mbox{stopping criterion}
\nonumber \\
2 N \hat{D}_{JS} &> & 8 + O(\frac{1}{N}) \nonumber \\
\mbox{BIC-based} & & \mbox{stopping criterion}  \nonumber \\
2 N \hat{D}_{JS} &> & 4 \log(N) + O(1) + O(\frac{1}{\sqrt{N}}) +O(\frac{1}{N}).
\end{eqnarray}
It is interesting to compare these criteria with those
in the hypothesis testing framework.  Setting the value of 8
to the $\chi^2_{df=4}$, the corresponding significance level
(p-value, tail-area) is  0.091578. In the hypothesis testing framework,
it is allowed to set an even more relaxed significance level such as
0.1, but in the AIC-based model selection, 0.091578 is the limit
of allowed levels. Similarly, with a given sequence length $N$,
the limit of allowed levels can be determined by a BIC-based model
selection; for example,  if $N=1$Mb, the significance level is
2.8631 $\times 10^{-11}$.  Again, the significance level can not
be more relaxed than these limits.

Besides limits on the relaxed side of the stopping criterion, there
are no theoretical limits on the stringent side. One model can be
``marginally better" than another model, ``moderately better",
or ``much better", etc. We will show that one can gradually make
the stopping criterion more stringent so that average domain
size is increased. For convenience, we define the ``strength"
of a 1-to-2 segmentation as the percentage increase of
$2 N \hat{D}_{JS}$ over the BIC-defined stopping threshold:
\begin{equation}
\label{eq_seg_strength}
strength = \frac{ 2 N \hat{D}_{JS} - 4 \log(N)}{ 4 \log(N) }.
\end{equation}
The strength has to be larger than 0, but it has no upper limit.

\section{Results}

Since the AIC-based stopping criterion is more relaxed than the
typical 0.01-significance-level test, one will end up with more
domains than from the program discussed in \cite{ber99}. The BIC-based
stopping criterion, however, is more interesting for our purpose,
because it provides a theoretical justification for using a much
more stringent stopping criterion than those typically used in
the hypothesis testing framework. We illustrate the BIC-based
segmentation by three DNA sequences with a wide range of
sequence lengths.

\subsection{Lambda phage} 
Fig.1 shows the result for $\lambda$ bacteriophage
($N=48,502$ b) \cite{lambda}.  This sequence has been tested with various
segmentation methods in \cite{braun}.  There are several pieces of
information displayed in Fig.1: the domain borders obtained by the
BIC-based segmentation on the original four-symbol sequence (upper bars);
the borders segmented by the two-symbol (CG vs. AT) sequence
(middle bars); the borders obtained by the AIC-based segmentations
(with higher-order terms included) (dots; due to the limitation of
resolution, individual dots can be hard to see); a moving-window C+G
content along the sequence; the strength of the segmentations as
defined in Eq.(\ref{eq_seg_strength}) (lower spikes); and the
sequential order of early-rounds of segmentations (e.g. the first
partition point from the 1-to-2 segmentation is labeled ``1st").
We note the following: (1) Segmentation results from the four-symbol
sequence and the two-symbol sequence are different. (2) The number
of domains by segmenting the four-symbol sequence is 6, which
is the same as results from a two-state hidden Markov segmentation
as discussed in \cite{braun}.  (3) Early-rounds of 1-to-2
segmentations are usually  the ``strongest" (with largest strengths).
(4) Even without any tuning of parameters (in contrast to tuning of
the significance level in the hypothesis testing framework), the
BIC-based segmentation manages to obtain a reasonable number of 
domains (AIC-based segmentation, on the other hand, leads to too
many domains).

\subsection{MHC} 
Fig. 2 shows the result for the human major histocompatibility
complex (MHC) sequence ($N=3,673,778$ b) \cite{mhc}. The MHC sequence
is a highly gene-rich region (with more than 200 genes) that is located
on the short-arm of chromosome 6 of the human genome. The segmentation
result captures the complexity of this sequence. With so many
domain borders in Fig.2, we only show those that have strengths
larger than 100\%.
Historically, the MHC sequence is divided into three domains
(in the telomere-to-centromere direction): class-I, class-III
(C+G-rich), class-II (C+G-poor). The MHC sequencing project
added another C+G-rich domain to the end: extended-class-II.
Interestingly, the border of these domains can be easily detected
by segmentation (these results are from the two-symbol segmentation):
I/III: i= 1,841,871, strength= 23679.6\%,
III/II: i= 2,483,966, strength= 17084.7\%,
II/extended-II: i=3,384,907, strength=28849\%.
These three 1-to-2 segmentations are the strongest. With
this segmentation result, the domain sizes of class
I, III, II, and extended-II are: 1.84 Mb, 0.64 Mb, 0.90 Mb
and 0.29 Mb.
The number of segmented domains in the MHC sequence is very large
(1260 from the BIC-based two-symbol segmentation and 1828 from the
BIC-based four-symbol segmentation). Segmentation with the minimum
requirement (i.e. for BIC to decrease) not only leads to large,
100kb-plus domains, but also leads to smaller-scaled base composition
fluctuation.  This ``domains-within-domains" phenomenon has been
discussed in \cite{wli94,ber96,wli97,complex2}. If one is
only interested in isochores, i.e., large DNA segments with usually
300 kb or longer that have relatively homogeneous base composition 
\cite{isochore,isochore2}, a more stringent criterion has to be 
used (to be discussed later).

\begin{figure*}
\begin{center}
  \begin{turn}{-90}
  \epsfig{file=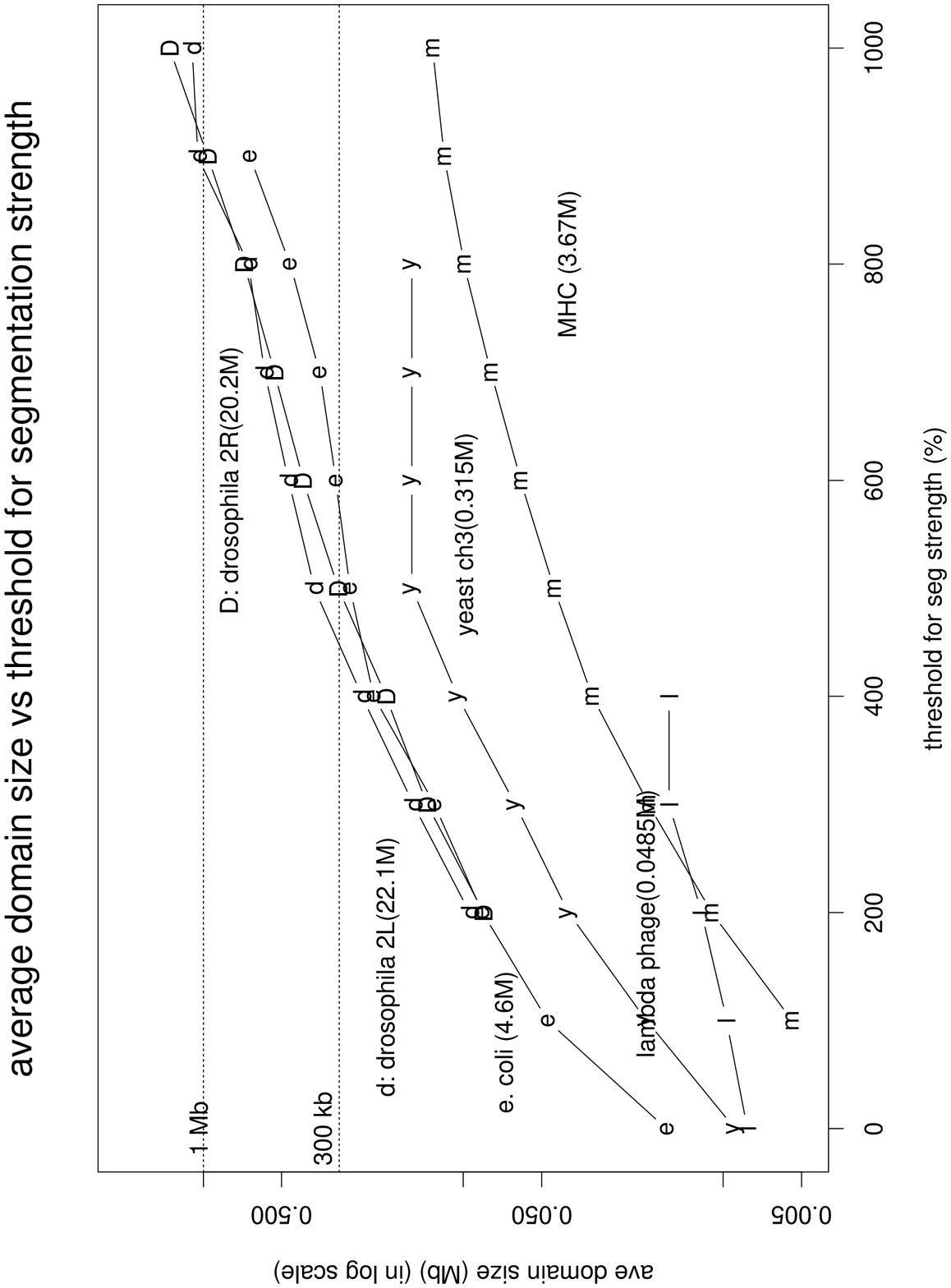, height=9cm, width=9cm}
  \end{turn}
\end{center}
\caption{
Average domain size vs. threshold for segmentation strength.
}
\label{fig4}
\end{figure*}

\subsection{Left-arm of Drosophila chromosome 2} 
The last sequence to
be segmented is the left arm of {\sl Drosophila melanogaster}
chromosome 2 ($N=22,075,671$ b)\cite{dro}. There is 1.78\% of
the sequence that is not determined (symbol ``$n$" or ``$N$").
To preserve the location information, these undetermined symbols
are replaced randomly by the four nucleotides (according to
the actual base composition of this sequence). Only the 1-to-2
segmentations with strength larger than 200\%  are included in
Fig.3, and only the result for the four-symbol sequence is
displayed. The segmentation of a four-symbol sequence is more
likely to cut the telomere (as well as centromere) at an earlier
stage than the corresponding two-symbol sequence; and this
is shown in Fig.3. This observation can be used to delineate
complex sequence patterns in telomere sequences (D Kessler and W Li,
in preparation).

Although the drosophila sequence is much longer than the MHC sequence,
there is only one 1-to-2 segmentation of the drosophila's
left-arm of chromosome 2 that has a similar strength as those of
the MHC sequence leading to domain borders.  This occurs
at position 6,959,803 with the strength 16768\%. If we use
a similar strength criterion as that used in delineating three
domain classes in the MHC sequence, there is only one domain
border in this sequence.

\subsection{How stringent the stopping criterion has to be to reach
a certain domain size} 
Since the model selection framework
only provides a limit on the relaxed end of the stopping criterion,
the stringent end is in principle open. Nevertheless, we can
empirically determine the typical domain size as a function of
the stringency of the stopping criterion. Fig.4 shows the
average domain sizes versus the threshold value for the strength,
all based on the four-symbol segmentation. Besides the three
sequences used in Figs.1-3, results from {\sl Escherichia coli}
($N=4,639,221$ b), the right arm of {\sl Drosophila melanogaster}
chromosome 2 ($N=20,228,487$ b), and yeast {\sl Saccharomyces cerevisiae}
chromosome 3 ($N=315,341$ b) are also included.

Fig.4 shows that for the drosophila sequence (not surprisingly, the
left and right arms of chromosome 2 behave similarly),
the segmented domains have an average size of 1 Mb when the
threshold of segmentation strength is set at 900\%. When the
threshold of strength is set at 500\%, the average size of
the segmented domains is around 300 kb -- a rough minimum
size for an isochore \cite{isochore}.
Fig.4 also shows that different sequences behave differently.
For the MHC sequence, for example, it is very difficult to
delineate only the larger domains. To reach the average size of
300 kB, the threshold of strength has to be set at 1300\%,
and to reach 1 Mb, the threshold has to be around 2800\% $\sim$
3200\%.  Another way to state this difference is that the MHC
is more ``complex" than other sequences in Fig.4, in the
sense of the existence of a huge number of domains.

Plots like Fig.4 are similar to the ``compositional complexity"
\cite{complex1,complex2,complex3}. The difference is that
in \cite{complex1,complex2,complex3}, not only the number of domains,
but also the base composition difference between domains is
part of the measure of complexity. In Fig.4, it is purely
the number of domains. Nevertheless, the plot of Fig.4 is useful
because it provides practical guidance on the choice of stopping
criterion at the stringent end. This choice will subjectively
depend on what length scales are of interest to the investigator.

\section{Acknowledgments}
The work is supported by the NIH grant K01HG00024.
The author acknowledges discussions/communications with
J\'{o}se Oliver, Pedro  Bernaola-Galv\'{a}n, Ivo Grosse,
Ken Burnham,  and Yaning Yang.

\end{document}